**Title:** Dynamical phenomena at the inner edge of the Keeler gap

Authors: Radwan Tajeddine[1], Philip D. Nicholson[2], Matthew S. Tiscareno[3], Matthew M. Hedman[4], Joseph A. Burns[2,5], Maryame El Moutamid[1]

[1]Center for Astrophysics and Planetary Science, Cornell University, Ithaca, NY 14853, USA
[2]Department of Astronomy, Cornell University, Ithaca, NY 14853, USA
[3]Carl Sagan Center for the Study of Life in the Universe, SETI Institute, 189 Bernardo Avenue #200, Mountain View, CA 94043, USA
[4]Department of Physics, University of Idaho, Moscow, ID 83844, USA
[5]College of Engineering, Cornell University, Ithaca, NY 14853 USA

**Abstract**

We analyze several thousand Cassini ISS images in order to study the inner edge of the Keeler gap in Saturn's outer A ring. We find strong evidence for an *m*=32 perturbation with a mean amplitude of radial variation of 4.5 km. Phase analysis yields a pattern speed consistent with the mean motion of Prometheus, indicating that this pattern is generated by the 32:31 Inner Lindblad resonance with Prometheus. In addition, we find evidence of 18-lobed and 20-lobed patterns with amplitudes of ~1.5 km. These patterns, whose rotation rates correspond to resonance locations ~4 km interior to the gap edge, are believed to be normal modes. The former is probably related to the nearby 18:17 (*m*=18) resonance with Pandora. In addition to these resonant and normal mode patterns, we also observe multiple localized features that appear to move at the local keplerian rate and that persist for only a few months. One hypothesis is that different groups of ring particles at the inner edge of the gap may be reacting differently to the resonance with Prometheus, with local variations in the forced eccentricity and/or pericenter; an alternative hypothesis is the existence of several unseen objects embedded at or near the inner edge of the Keeler gap, similar to those suspected to exist at the outer edges of the A and B rings (Spitale and Porco 2009, 2010). In either case, observations of the ring edge at opposite ansae demonstrate that the localized features must be on eccentric orbits.

## 1. Introduction

Among the many curious features that characterize the rings of Saturn is a ~37 km wide sharp-edged gap in the outer A ring known as the Keeler gap. In an early study of the Keeler gap using Voyager images, Cooke (1991) showed that the local gap width varied from ~35 to ~40 km, and that at least part of this variation could plausibly be attributed to perturbations by the Prometheus resonance, described below. Cooke also found evidence for a shorter-wavelength perturbation, possibly a 50-lobed pattern. However, given the small number of Voyager images with sufficient resolution for such a study, Cooke (1991) was unable to derive a single, self-consistent model for the Keeler gap edges. To date, no analysis of the long-wavelength variations in the edges of the Keeler gap using the much more extensive set of Cassini imaging data has been published.

Porco et al. (2005a) suggested, based on high-frequency radial perturbations seen at the gap edges, that a small satellite orbited within the gap, a prediction that was soon confirmed by additional Cassini images (Porco et al 2005b). This satellite, now known as Daphnis, is believed to maintain the gap via the same shepherding mechanism originally proposed to maintain the narrow uranian rings by Goldreich & Tremaine (1979). Torques exerted on the ring by the satellite prevent the gap edges from closing due to viscous spreading. In a visible manifestation of this process, as ring particles reach Daphnis' longitude, their orbits are gravitationally disturbed and they acquire free eccentricities and inclinations (the latter due to Daphnis' inclined orbit), which produce prominent radial and vertical wavy structures on both edges of the gap (see Fig. 1). These are discussed by Tiscareno et al. (2005), Weiss et al. (2009), and Torrey et al. (2010). Unlike the similar perturbations produced by the satellite Pan at the edges of the Encke Gap (Cuzzi & Scargle 1986), however, the wavy features on the Keeler gap edges damp quite rapidly, within few degrees in longitude.

Another type of structure present only at the outer edge of the gap are small-scale (a few hundred-meters radially by ~1° in azimuth) features that extend radially into the gap from the outer edge (Porco et al. 2005). These so called "wisps" orbit Saturn at the local Keplerian rate, and are believed to be due to local perturbations created by small objects embedded in the ring near the outer edge of the gap (Arnault & Tiscareno, 2016). Wisps appear to be unique to the outer edge of the Keeler gap.

The inner edge of the gap (i.e., the edge closer to Saturn) is the subject of this paper. Its mean radius, based on fits to over 100 stellar and radio occultation profiles (French et al. 2017), is 136,484.8±0.3 km, and it coincides closely with the 32:31 Inner Lindblad Resonance (ILR) with Prometheus. The resonance location of 136482.4 km, is only ~2 km interior to the edge of the gap. In general, the mean motion of a ring particle in a first-order ILR with a satellite obeys the relation

$$(m-1)n + \varpi - mn_s = 0, \tag{1}$$

where, $n$ and $n_s$ are the mean motions of the ring particle and satellite, respectively, $m$ is the azimuthal wave number and $\varpi$ is the local apsidal precession rate due to Saturn's gravity field (see Eq. 5 below). After $m$ orbits such a ring particle will return to the same longitude relative to the satellite, at the same mean anomaly, while the latter would have orbited Saturn $m$-1 times. For higher-order resonances, $n_s$ is replaced by the pattern speed $\Omega_p$, the angular rotation rate of the external perturbation. Eq. (1) also formally describes an Outer Lindblad Resonance (OLR) if we adopt the convention that $m$ is negative for such resonances (Hedman et al., 2013). For an ILR, $\Omega_p < n$ whereas for an OLR, $\Omega_p > n$.

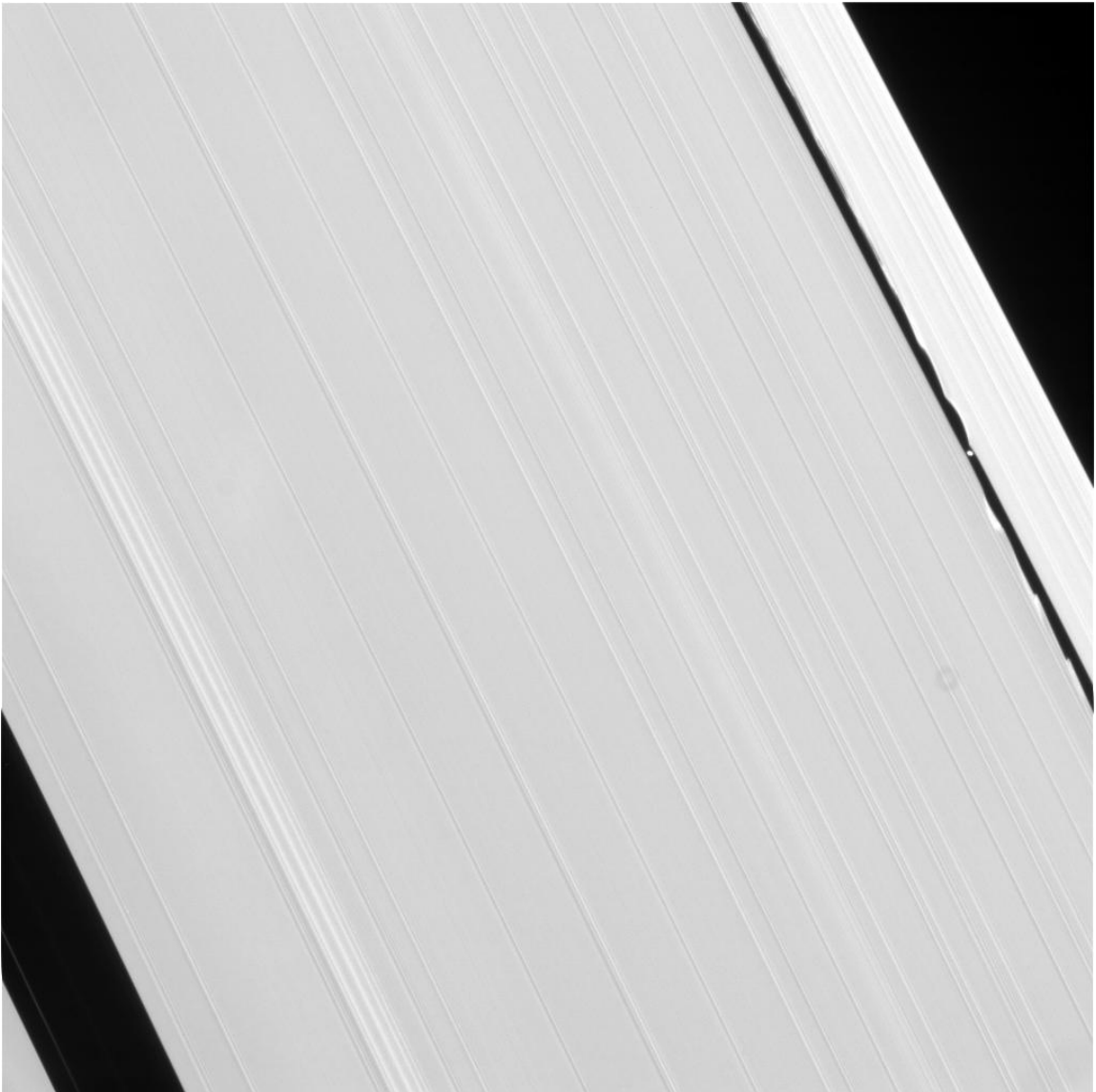

**Figure 1.** A Cassini image of the outer part of the A ring (Image name: N1540685777). From left to right are the Encke gap, the Keeler gap (with Daphnis in the middle causing the waves on the gap edges), and the outer edge of the A ring. The linear features in the ring are density and bending waves created by resonances with satellites exterior to the rings.

At Lindblad resonances, perturbations from the satellite excite the ring particles' eccentricities. As a result, in the reference frame of the perturbing satellite, the ring edge develops an $m$-lobed pattern, representing the $m$ apoapses and periapses that each ring particle goes through before returning to the same longitude it started from in the satellite's frame. In the case of the inner edge of the Keeler gap, the satellite is Prometheus and $m=32$. The radius of the inner edge of the Keeler gap is thus expected to vary as a function of time and inertial longitude as

$$r(\lambda,t) = a + ae\cos\left[m(\lambda - \Omega_p t - \lambda_0) + \delta\right] \qquad (2)$$

where $a$ is the average radius of the ring edge, $e$ is its eccentricity, $\lambda$ is the inertial longitude of the ring edge, $\Omega_p$ is the rotation rate of the $m$-lobed pattern (equal in this case to the mean motion of Prometheus $n_s$), $t$ is the elapsed time since our adopted epoch of J2000, $\lambda_0$ is the inertial longitude of Prometheus at $t=0$, and $\delta$ is a phase angle. Theoretically, $\delta$ should be either 0 or 180 degrees, meaning that either an apoapsis or a periapsis of the pattern, respectively, should be aligned with the satellite's longitude. We follow here the notation of Nicholson et al. (2014a,b), except for the definition of the phase angle; in terms of their phase angle $\delta_m$, we have $\delta = m(\lambda_0 - \delta_m)$.

In this work we investigate phenomena at the inner edge of the Keeler gap by analyzing 58 FMOVIE mosaics based on several thousands of Cassini images taken between 2005 and 2014. In section 2 and 3 we describe the observations used in this work, and then present evidence of the anticipated $m=32$ perturbation at the inner edge of the gap. We also show that the resonant perturbation due to Prometheus is not the only distortion present, and in section 4 we present evidence for the presence of both $m=18$ and $m=20$ normal modes at the gap edge. In section 5, we discuss additional perturbations at the inner edge of the Keeler gap that appear to be unrelated to the resonant and normal mode signatures. In section 6, we discuss the viscosity of the ring and the torques exerted on the gap edges. Our results are summarized and discussed in section 7.

## 2. Analysis of FMOVIE mosaics

Among the many different types of observations of the rings by Cassini ISS is a group known as FMOVIEs. These are sequences of a large number (a few tens to hundreds, see Table 1) of images that were specifically designed to track perturbations of Saturn's F ring by Prometheus and various ring-embedded objects (Murray et al. 2005; 2008). Fortunately, the Keeler gap appears in most of those observations, a circumstance that allowed us to carry out this study. A typical FMOVIE has a duration of 10-15 hours, comparable to the 14.9-hour orbital period of the F ring and the 14.3-hour period of objects in or near the Keeler gap.

Since it is impossible for Cassini to take a single, high-resolution, instantaneous snapshot of the whole F ring, multiple images must be put together to build mosaics of the ring. There are two types of FMOVIE observations. The most common are the "Stare" variety, where the ISS Narrow Angle Camera (NAC; Porco et al. 2004) is pointed towards one fixed inertial longitude and observes the ring particles passing through the camera's frame as they orbit Saturn. This provides a good view of rotating patterns that shape the ring azimuthally, but is insensitive to any global eccentricity or inclination, since such patterns do not precess appreciably in the course of a single observation, and so do not cause any change in the observed radius of the ring. Therefore, if the ring edge had an eccentricity or an inclination, its mean radius would be constantly smaller or larger than the mean one depending on its orbital phase when the edge was observed; this does not affect our results here because here we are studying only high-frequency radial variations. A less frequent type of observation is one that follows one segment of the ring as its particles orbit Saturn, intended to show how one particular group of ring particles traces out their eccentric orbits. In this study, we use only the first type.

*Cassini Image processing: methods*

In order to analyze a Cassini image, some starting information is required. The label of every image contains the observation time, which allows us (with the help of the tools in the NAIF SPICE library: http://naif.jpl.nasa.gov/naif/; Acton 1996) to obtain information on the instantaneous position of the spacecraft and the pointing vector of the ISS NAC. The former is known to a precision of ~1 km, which is reflected as uncertainty of ~1km on the position of an observed object. The latter is less well known a priori: the projected error in an image is on the order of tens of pixels, yielding a typical error at the Keeler gap of 50-100 kilometers. The most common way to correct this error is with star astrometry (Tajeddine et al. 2013; 2015; Cooper et al. 2014). However, this method cannot be employed for most of the images used in this study, as the rings cover a large proportion of the image, reducing the chance of detecting many stars and so correcting the errors in the camera pointing. Another method of pointing correction is to use a circular ring feature, such as the outer edge of the Keeler gap. The absence of any resonant perturbation (it still has minor perturbations such as Daphnis waves and wisps) makes it a circular edge, within the limit of Cassini ISS resolution, with a radius determined by occultation data of 136522.18±0.09 km (French et al. 2017). This edge can thus be used as a fiducial to correct the camera pointing (discussed further below). After testing both methods, we noticed that the pointing correction based on stellar astrometry created in some cases unaligned trends when building edge profile mosaics. The reason is that a pointing correction using catalogue stars does not make up for any error in the spacecraft position; but when using the outer edge of the Keeler gap both the camera pointing and spacecraft position are corrected. Therefore, we adopted the second method to navigate our Cassini images.

To correct the pointing using the outer edge of the Keeler gap, we first locate the edge by calculating the derivative of the brightness curve as a function of pixel number along a line going from the dark gap to the bright ring (this is similar to the method used in satellite limb measurements; Tajeddine et al. 2013). By fitting a Gaussian function to the peak of the brightness derivative curve, we obtain the edge position with an uncertainty on the order of 0.1 pixels. The camera pointing for each image is then corrected by minimizing the distance between the measured edge and the predicted one projected on the image via SPICE geometry calculations. Once the camera pointing is corrected, we use the SPICE library to determine the ring radius and longitude coordinates for each image pixel.

Finally, to build an edge mosaic of the Keeler gap, we must reproject the relevant portion of each image into (inertial longitude, radius) coordinates. We then measure the edge profile in the reprojected image, applying the same method used on the outer edge to correct the pointing as described above. The original inertial longitudes are then converted to co-rotating longitudes via the expression $\theta=\lambda-\lambda_s$, where $\lambda$ is the measured inertial longitude and $\lambda_s$ is the instantaneous inertial longitude of a satellite, obtained from the SPICE ephemeris (if we built the mosaic in the inertial frame, all the images and edge profiles would be overlapping, because the Cassini camera usually stares at one inertial longitude during an FMOVIE sequence). The mosaic is then presented in radius and co-rotating longitude ($\theta$) coordinates.

## 3. The Prometheus Perturbation

In Figure 2 we present a mosaic of the Keeler gap, made by assembling 130 NAC images from an FMOVIE obtained by Cassini ISS in 2013. This reprojected mosaic shows the radial distance from Saturn in kilometers as a function of longitude from Prometheus. As expected, the inner edge exhibits clear quasi-sinusoidal variations in radius, which form a somewhat irregular 32-lobed pattern in Prometheus' frame. The outer edge shows no evidence of any major perturbations in radius, but this is because we assumed it to be circular in order to navigate the images.

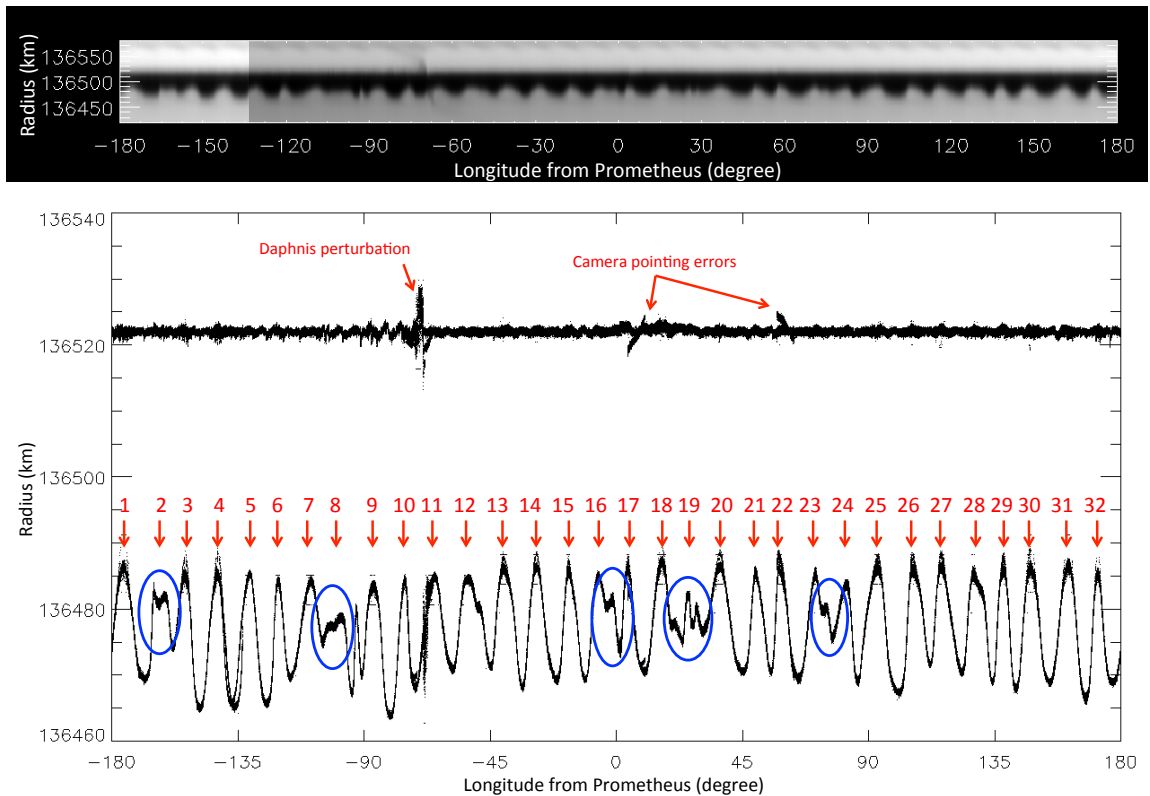

Figure 2. A 360 degree mosaic of the Keeler gap (constructed in Prometheus' frame) made from observations taken by Cassini ISS on 21 August 2013, on its 196[th] orbit (see Table 1 for further details). The top panel shows the reprojected mosaic of actual images while the bottom panel shows the edge profiles extracted from the mosaic. The disturbances around -70° are the Daphnis waves, and the glitches at ~5° and ~50° are due to an error in the camera pointing correction for two images. Numbered arrows highlight the 32-lobed pattern on the inner edge. The circles show some of the features that do not fit the dominant $m=32$ pattern. The brightness cut at ~ -135° (top panel) represents the beginning/end of an observations sequence, thus, the first and last image of the sequence

are separated by ~15 h where the spacecraft has changed its viewing angle relative to the Sun and the rings, which explains the brightness difference.

Although the *m=32* pattern seems to be the dominant perturbation at the inner edge of the gap, as anticipated above, irregularities in this pattern raise the possibility of additional perturbations, resonant or otherwise. Variations in the amplitude of the pattern suggest the possibility of beating with another periodic signal. We also draw the reader's attention to the "wiggles" at −163, −94.5, −3.5, +23 and +73.5°, which do not seem to fit the dominant *m=32* pattern.

In order to investigate the nature of these irregularities, we have analyzed similar edge profiles obtained by Cassini ISS from a total of 58 MOVIEs between 2005 and 2014. These data sets are listed in Table 1, along with their durations, inertial longitudes, radial resolutions, and the number of images per movie. For each edge profile, we performed a Fourier analysis, treating the corotating longitude *θ* (this time in the local keplerian frame, with $\lambda_0 + \Omega_p t$ in Eq. (2) is replaced by $n(t - t_0)$ ) as the independent variable and radius as the dependent variable; the resulting spatial frequencies were converted to the equivalent value of *m*, representing the number of lobes per 360° of longitude in the corotating frame.

Figure 3 shows the mean amplitude spectrum averaged over the individual spectra obtained from the Fourier transforms of the 58 edge profiles; the abscissa is labeled as *|m-1|* because any *m*-lobed resonant perturbations would appear with *m*-1 lobes in the local keplerian frame. We adopted this approach because, while a mosaic constructed in the reference frame of the perturbing satellite (such as that in Fig. 2) will show *m* lobes, each *potential* perturbation has its own unique pattern speed, given by Eq. (1). Resonant perturbations due to another satellite, for example, would be associated with a different pattern speed and would thus not show an integer numbers of lobes in such a mosaic. It can be shown, however (see Eq. (9) of El Moutamid et al. 2016) that any perturbation which obeys Eq. (1) (ie., an inner or outer Lindblad resonance) will lead to a signature with *|m-1|* lobes if plotted in a frame rotating at the local keplerian rate, *n*.

As expected from Fig. 2, the strongest signal in Fig. 3 is at *|m-1|* = 31 with a mean amplitude of about 4.5 km. This is consistent with an ILR-type perturbation with *m=32*. Weaker signals appear at *|m-1|* = 17 and 19, and are discussed in the next section. We do not find any evidence for a signature at *m=50* as reported by Cooke (1991), based on a limited number of Voyager images. In hindsight, this signal could have been due to a chance observation of those parts of the ring where the dominant pattern consists of irregular high frequency features such as those seen around -95° or +23° in Fig. 2.

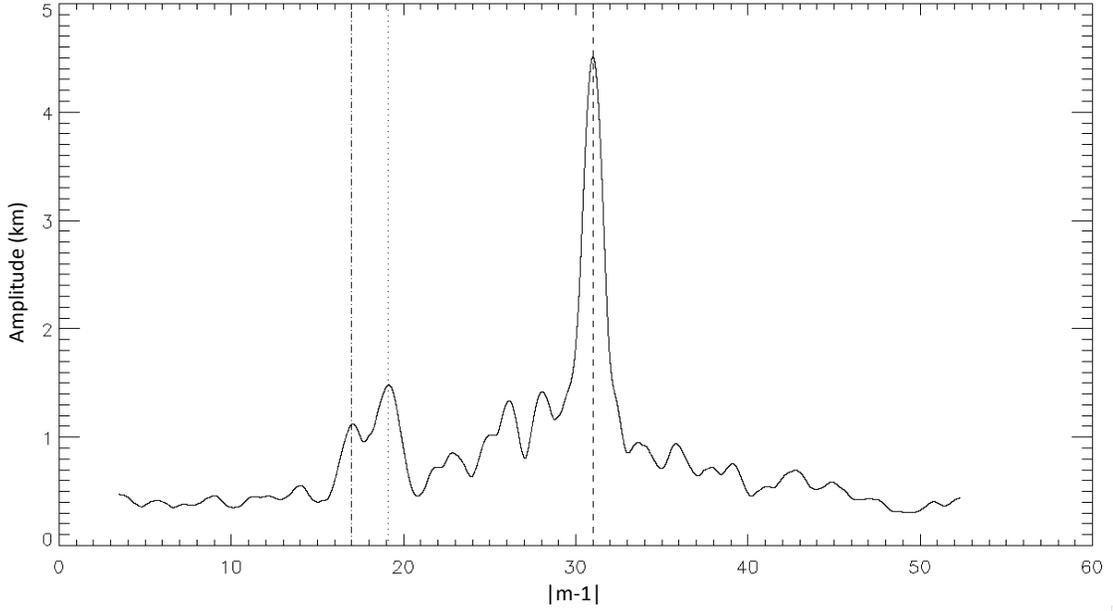

Figure 3. Averaged amplitude spectrum obtained from the Fourier transforms of profiles extracted from the 58 mosaics of the inner edge of the Keeler gap listed in Table 1. The spatial frequency has been converted to |$m$-1|, as the profiles were analyzed in the local keplerian frame (see text).

To investigate whether this signal is indeed caused by the resonance with Prometheus, we performed a phase analysis by fitting Eq. (2) to the edge profiles measured in individual FMOVIE sequences. We set $m$=32 and the pattern speed to $\Omega_p$=587.285°/day, equal to the mean motion of Prometheus. (This value was obtained by averaging the mean motion derived from the numerically-integrated ephemeris provided in the SPICE software package during the period from 2004 to 2015. Note that it is slightly different from the value given by the JPL Horizons, website: http://ssd.jpl.nasa.gov/horizons.cgi). Each image is associated with one observation time $t$, but up to a thousand edge points, each with their own inertial longitude $\lambda$; from these we computethe model radius, $r_c(\lambda,t)$ and minimize the $\chi^2$ statistic for each FMOVIE by fitting for the amplitude $ae$ and the phase $\delta$ of the perturbation. We use a simple, unweighted measure given by

$$\chi^2 = \frac{\sum_{i=1}^{N}\left[r_o(\lambda_i,t_i) - r_c(\lambda_i,t_i)\right]^2}{N} \quad (3)$$

where the indices $o$ and $c$ represent the observed and calculated radii, respectively, and $N$ is the total number of points used in the fit, which is on the order of 20,000 data points per FMOVIE.

For individual FMOVIES, the fitted amplitudes range from 2.5 to 6 km, with an average value of 4.5 km. The fitted phases are shown in Fig. 4; each point represents a fit to a single FMOVIE. The phase varies between 150° and 190°, with a mean value of 166° and a **standard deviation of 14°.** The concentration of phases near 180° means that the $m$=32 pattern is following Prometheus' motion with a pericenter that is nearly aligned in

longitude with Prometheus. However, the pattern leads the satellite by an average of 14°
(Note that this is the phase lag, as defined in Eq. (2). In order to convert this to a lag in
longitude, it must by divided by *m*=32, implying a lag of Δ=0.4° in one of the minima in
the pattern with respect to Prometheus' longitude). A similar lag was observed in the
*m*=2 pattern at the edge of the B-ring, which is forced by the 2:1 inner Lindblad
resonance with Mimas (Spitale & Porco 2010), and interpreted by these authors in terms
of the ring's effective viscosity *v*. We will return to this possibility and its implications
for torque balance at the gap edges in Section 6 below.

Finally, we note that there appear to be small but systematic variations in the phase over
the 9-year period of our observations, with δ increasing prior to 2010 and perhaps
decreasing after 2012. Without a longer span of data, it is impossible to tell if this
represents an oscillation (or libration) in δ about its long-term mean value or simply
errors in the phase determinations, or maybe a long-period longitude variation in
Prometheus' orbit.

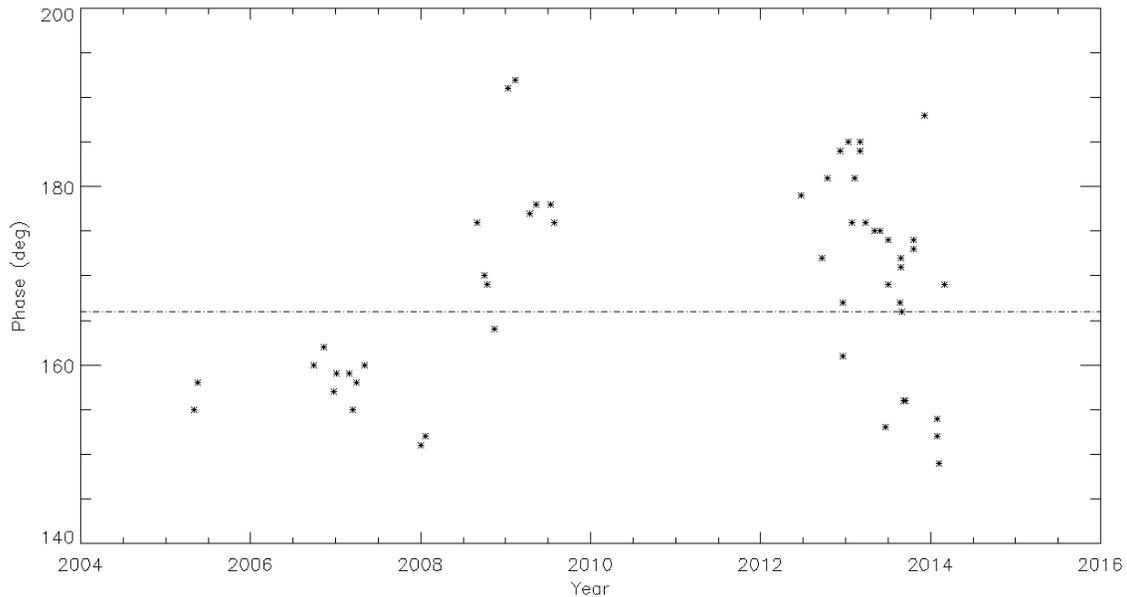

**Figure 4.** Fitted phase *δ* (cf. Eq. 2) for each of the mosaic profiles of the inner edge of the
Keeler gap, assuming *m*=32 with a pattern speed equal to Prometheus' mean motion.
Cassini was in a near-equatorial orbit in 2006 and 2010-12, resulting in the absence of
ring observations in these periods. The dashed line represents the mean value of 166° ±
14°.

**4. Weaker perturbations**

As seen in the previous section, the average amplitude spectrum in Fig. 3 shows strong
evidence for the effects of the 32:31 Prometheus ILR at the inner edge of the Keeler gap.
However, this figure also shows two additional noteworthy peaks at |*m*-1| = 17 and 19,
each of which has two possible interpretations. If they are due to inner Lindblad
resonances with *m*=18 and *m*=20, then we might expect to find a nearby 18:17 and/or
20:19 resonance with an exterior satellite. On the other hand, if they are due to outer

Lindblad resonances with $m=-16$ or $m=-18$, we must look for a 16:17 and/or 18:19 resonance with an interior satellite, or perhaps with Saturn itself. (For further discussion of this ambiguity, the reader is referred to Section 2 of El Moutamid et al. (2016).)

In order to check these possibilities, we calculated the radii for all first-order satellite resonances in this region, using Eq. (1) and appropriate expressions for the mean motion $n$ and the apsidal precession rate $\dot{\varpi}$ (Nicholson & Porco, 1988; French et al., 1982, Borderies-Rappaport & Longaretti 1994):

$$n \approx \sqrt{\frac{GM}{a^3}} \left\{ \begin{array}{l} 1 + \frac{3}{4}J_2\left(\frac{R}{a}\right)^2 (1+4e^2-16\sin^2 i) - \left[\frac{15}{16}J_4 + \frac{9}{32}J_2^2\right]\left(\frac{R}{a}\right)^4 \\ + \left[\frac{27}{128}J_2^3 + \frac{45}{64}J_2 J_4 + \frac{35}{32}J_6\right]\left(\frac{R}{a}\right)^6 \end{array} \right\}, \quad (4)$$

$$\dot{\varpi} = \sqrt{\frac{GM}{a^3}} \left\{ \frac{3}{2}J_2\left(\frac{R}{a}\right)^2 (1+e^2-2\sin^2 i) - \frac{15}{4}J_4\left(\frac{R}{a}\right)^4 + \left[\frac{27}{64}J_2^3 - \frac{45}{32}J_2 J_4 + \frac{105}{16}J_6\right]\left(\frac{R}{a}\right)^6 \right\} \quad (5)$$

where $M$, $R$, and $J_n$ are the mass, equatorial radius and zonal gravity harmonics of Saturn, (Jacobson et al., 2006) and $i$ and $e$ are the inclination and eccentricity of the ring particle orbit, respectively. (The original formulae given by Nicholson et al. (1988) contain an additional term that includes perturbations from external satellites, but these are negligible here and have been ignored for the purpose of this study.) In these expressions the quantity $a$ is the local mean (or epicyclic) radius, as defined by Borderies-Rappaport and Longaretti (1994), rather than the osculating semi-major axis.

We find that the closest resonance to the inner edge of the Keeler gap, besides the Prometheus 32:31 ILR, is the 18:17 ILR with Pandora located at 136,456.5 km, or ~28 km interior to the gap's inner edge. Inspection of stellar occultation profiles for this region shows a substantial density wave driven by this resonance that propagates as far as the gap edge before it is damped. This resonance is thus a prime candidate to explain the $m=18$ signature in Fig. 3.

*m=18 signature*

To test whether the 18:17 ILR with Pandora is indeed responsible for the observed signal at $|m-1|=17$ in Fig. 3 (perhaps via the intermediary of the density wave), we carried out a similar phase analysis to that done for the Prometheus resonance in the previous section. After first subtracting the best-fitting $m=32$ signal from each edge profile, we fit the amplitude $ae$ and the phase $\delta$ in Eq. (2) to the residual perturbations, but this time setting $m=18$ and the pattern speed $\Omega_p$ equal to the mean motion of Pandora of 572.790°/day. In most cases we obtain a reasonable fit with an average amplitude $ae$ ~1.3 km as suggested by the spectrum in Fig. 3. However, the fitted phases are not constant, but show a systematically increasing trend, as shown in Fig. 5. (Here, we have 'unwrapped' the fitted phases to make the trend more obvious to the eye.) Usually, such a linear trend in phase signifies that the assumed pattern speed is incorrect, with the fitted phases making up for the error. Despite this problem, the tight grouping of the fitted phases about a straight line strongly suggests that the *m*=18 signal is real; otherwise the phases would have been distributed randomly between 0 and 360°.

The slope of the trend line in Fig. 5 suggests a correction to the assumed pattern speed of −0.155°/day (the slope must be divided by 18, and the sign is reversed because of the way we define the phase in Eq. (2), so that the correction is negative). Such a correction is much larger than the maximum variations in the mean motion of Pandora, which are of order 0.03°/day as seen in its numerically-integrated orbit. But the proximity of the density wave to the gap edge suggests an alternative model. If instead of using the mean motion of Pandora, we adopt the corrected pattern speed of $\Omega_p$=572.639°/day and recalculate the resonant radius using Eqns. (1, 4, and 5), then we obtain a resonant radius of 136,480.2 km, which is only ~4 km interior to the gap edge.

This suggests that the *m*=18 perturbation may instead be a *normal mode at the gap's inner edge,* similar to those reported by Nicholson et al. (2014b) and French et al. (2016) on numerous gap and ringlet edges in the C ring and Cassini Division. Such modes are essentially a free instability at a sharp ring edge, and can be thought of in terms of a pair of inward- and outward-propagating density waves trapped between a virtual Lindblad resonance and the gap (or ringlet) edge. Spitale and Porco (2010) have referred to this phenomenon as a 'resonant cavity', which seems an apt analogy. At an inner gap edge, only ILR-type modes are expected (cf. discussion in Section 2 of Nicholson et al. 2014b), such as we see here. But why an *m*=18 mode, rather than any other *m*-value? In this case, we hypothesize that the *m*=18 mode is triggered by the radial disturbance associated with the Pandora 18:17 density wave, whose pattern speed differs only slightly from the local frequency of a free *m*=18 mode.

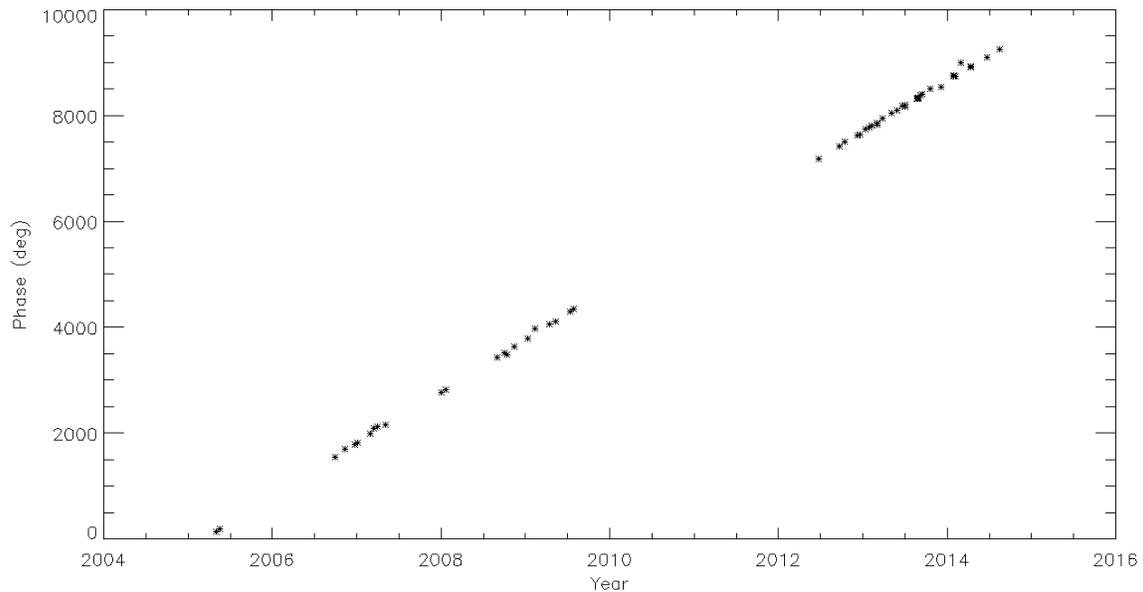

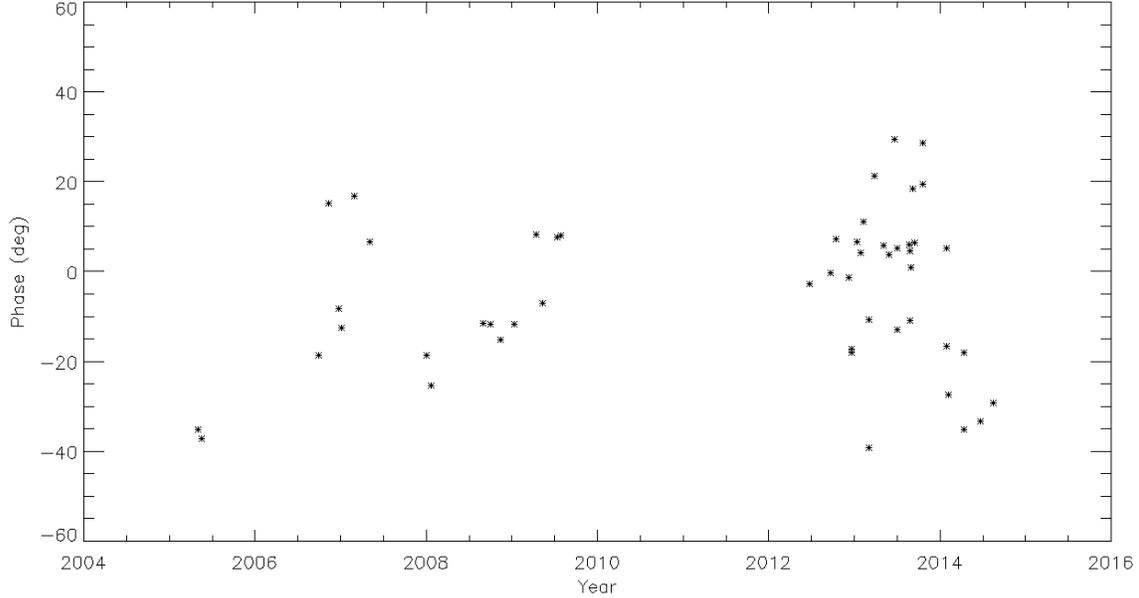

**Figure 5.** (Top Panel) Fitted phase $\delta$ (Eq. 2) for each of the mosaic profiles of the inner edge of the Keeler gap, assuming $m=18$ and a pattern speed equal to Pandora's mean motion, and removing the $m=32$ pattern. The original fitted phases have been unwrapped by eye to produce a monotonic sequence. (Bottom panel) A plot of the corrected phases after taking out the slope.

*m=20 signature*

Unlike the $m=18$ perturbation, we have not identified any external source for a 20:19 ILR in the vicinity of the Keeler gap. Such a hypothetical satellite would be located between Pandora and the F ring, and would surely have been discovered already. And a satellite responsible for an 18:19 OLR – which could also lead to a signature with $|m-1| = 19$ – would be located in the inner A ring, a hypothesis we can also safely rule out. So again we turn to the possibility of an ILR-type normal mode, this time with $m=20$.

We use Eq. (1) to calculate the pattern speed $n_s=\Omega_p$, corresponding to such a mode, using the local mean motion $n$ and apsidal precession rate $\varpi$ calculated from Eqns. (4) and (5), respectively. We find a pattern speed for $m=20$ at the inner edge of the Keeler gap of 575.961°/day. Using this pattern speed and $m=20$, we fit the residual perturbations in each edge mosaic by adjusting the amplitude $ae$ and phase $\delta$ in Eq. (2). Again the fits are mostly quite reasonable, with average amplitudes $ae$ ~1.7 km, consistent with the spectrum in Fig. 3. Figure 6 shows the fitted phases, which in this case show a slow negative drift over time. The mean slope suggests a correction to the assumed pattern speed of +0.027°/day (see previous section for the positive sign), resulting in a corrected pattern speed of 575.988°/day. Once again, the small scatter about the linear drift in the phases strongly suggests that the signal is real. Using the corrected pattern speed with $m=20$, we calculate the resonant radius using Eqns. (1, 4, and 5), again obtaining a radius of 136480.5 km, essentially the same as that of the *m*=18 mode. In this case, we can think of no particular reason that an *m*=20 normal mode might be excited.

It is also unclear why the *m*=20 normal mode should have almost the same resonant radius as that of the *m*=18 mode, but the similarity in *m*-values does imply that the resonant cavity widths should be quite similar (Nicholson et al. 2014a), so perhaps this is not surprising.

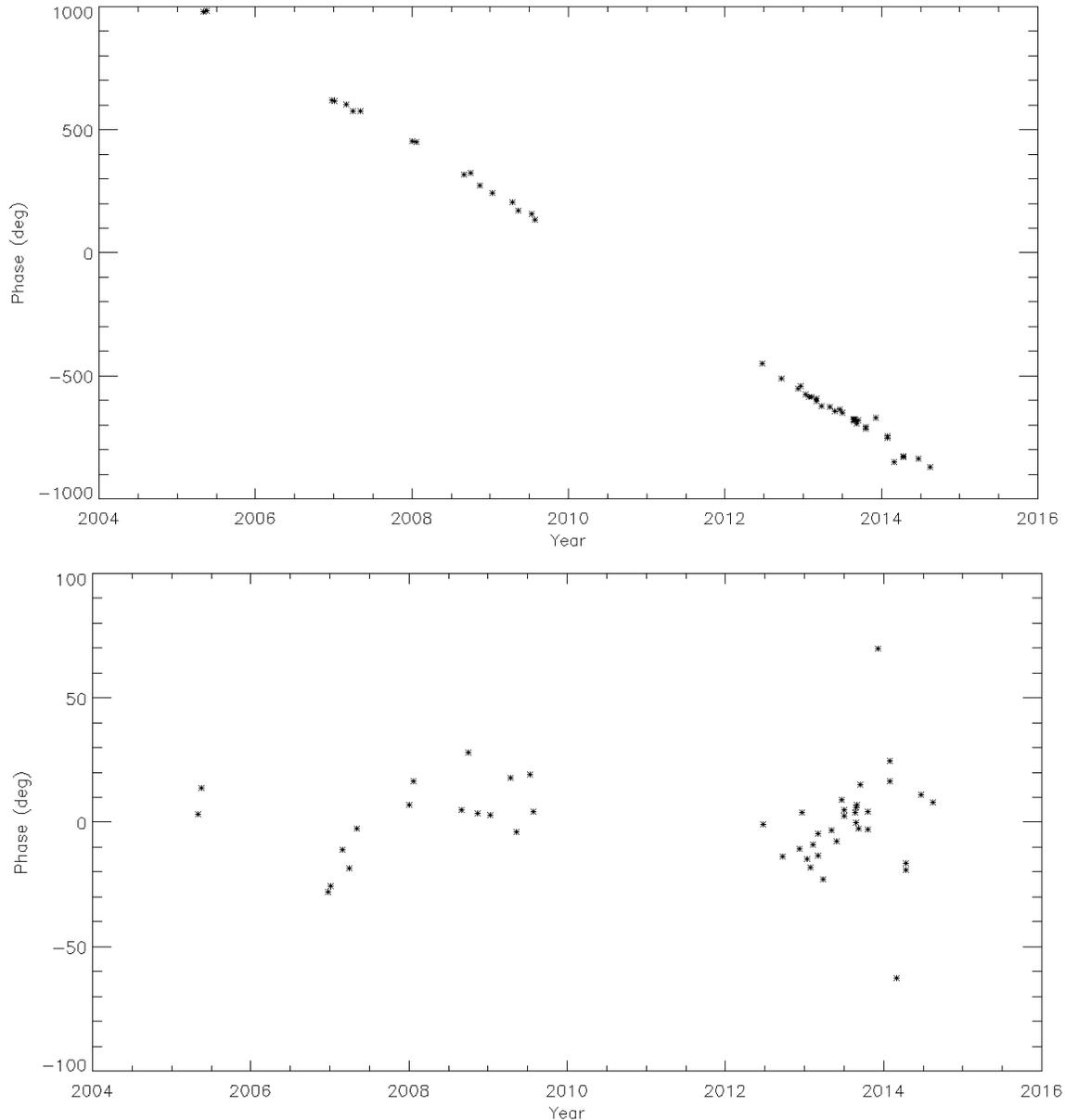

**Figure 6.** (Top panel) Fitted phase $\delta$ (Eq. 2) for each of the mosaic profiles of the inner edge of the Keeler gap, assuming *m*=20 and a pattern speed equal to 575.961°/day, and removing the *m*=32 pattern. As in Fig. 5, the original fitted phases have been unwrapped by eye to produce a monotonic sequence. (Bottom panel) A plot of the corrected phases after taking out the slope.

Figure 3 shows the possibility of existence of two additional patterns at $|m − 1| = 26$ and 28. However, phase analysis for those two $m$ numbers did not reveal any consistency between the phase fits from different movies; those features are most likely not real signals and may instead be due to aliasing from the $|m − 1| = 31$, 19, and 17 signals.

## 5. Other Phenomena

We turn now to the small-scale irregularities, or "wiggles", in the dominant $m=32$ pattern, such as those seen at −163, −94.5, −3.5, +23 and +73.5° in Fig. 2. Such features were first noted by Tiscareno et al. (2005), but without explanation. In their analysis of the edge of the B-ring, Spitale and Porco (2010) noticed similar irregularities in addition to the resonant perturbation due to the 2:1 ILR with Mimas and several normal modes. They interpreted these irregularities as being due to embedded objects orbiting near the edge of the ring and causing local distortions in the distribution of ring particle streamlines, analogous to the 'propellers' seen within the outer A ring (Tiscareno et al. 2006; 2008; Sremčević et al. 2007) or to the disturbance caused by the object unofficially named 'Peggy' at the outer edge of the A ring (Murray et al. 2014). A key argument in favor of this model was that the irregularities were observed to move through the resonant patterns at the local keplerian mean motion.

An ideal opportunity to test such a model for the high-frequency wiggles on the edge of the Keeler gap is provided by four sets of observations obtained by Cassini on rev 196, over a period of 1 week in August 2013 (listed as 196_3, 196_4, 196_5, and 196_6 in Table 1). The first of these observations is that illustrated in Fig. 2; the other three followed at intervals of 3.5, 2.0 and 1.9 days, respectively. In Figure 7 we compare profiles extracted from all four of these FMOVIEs, generated in a reference frame moving with the mean motion of Prometheus (i.e., $\Omega_p = n_s$). Although the $m=32$ pattern generally aligns from one observation to the next, the small-scale irregularities or deviations from this pattern clearly do not. However, closer inspection shows that the same features can be seen in each mosaic but that they drift systematically in longitude relative to Prometheus. For example, the sharp extra minimum at -94.5° in the first mosaic appears at -38° in the second, at +7° in the third and at +41° in the fourth. Over a period of 7.4 days this feature moves by +135.5°, or an average of 18.3 °/day; noting that the difference between the local keplerian rate of $n= 606.116°$/day and that of Prometheus is 18.69 °/day, we see that the offsets are approximately what would be expected for a feature moving at the local keplerian rate.

We can test this hypothesis by redoing the mosaics in the local keplerian frame, as shown in Fig. 8 (i.e., the longitudes are now measured relative to an arbitrary ring particle moving at the keplerian rate). An astonishingly good alignment appears in this figure, despite the fact that the ring particles have orbited Saturn more than 12 times during this period. Indeed, it appears from this example that almost all the small-scale structures in the edge profile are moving at the local keplerian rate. Note that this does *not* mean that the $m=32$ pattern itself is moving at the local Keplerian rate; on the contrary it clearly follows the mean motion of Prometheus, as shown in Section 2. But all the additional perturbations (that give each lobe its own distinctive shape) are moving at the keplerian rate. Thus, during each orbit around Saturn, these small-scale features move from one lobe in the $m=32$ pattern to the next. For example, the feature at -94.5° in the first mosaic in Fig. 7 moves ahead by five lobes in the second, by another four lobes in the third and by three more in the fourth, for a total of 12 lobes or 135° in longitude relative to the underlying $m=32$ pattern. The predicted shifts, given the intervals between the start times of the four observations, are 5.9, 3.4 and 3.2 lobes, respectively. (The actual shifts, noted above, differ slightly because the four movies began at different locations in the $m=32$ pattern. Close inspection of Fig. 7 will reveal the overlap region in each profile.)

In addition to the feature mentioned above, which we label as feature 'b' in Fig. 8, similar but less extreme examples are labeled as 'a', 'c', 'd' and 'e'. In most cases these take the form of `extra minima' between the resonant lobes, but feature 'd' is more complex and seems to involve two expected minima being split into four weaker minima. More careful scrutiny reveals that many of the other 32 minima are somewhat irregularly spaced, with spacings varying from as little as 8.5° to as much as 14°.

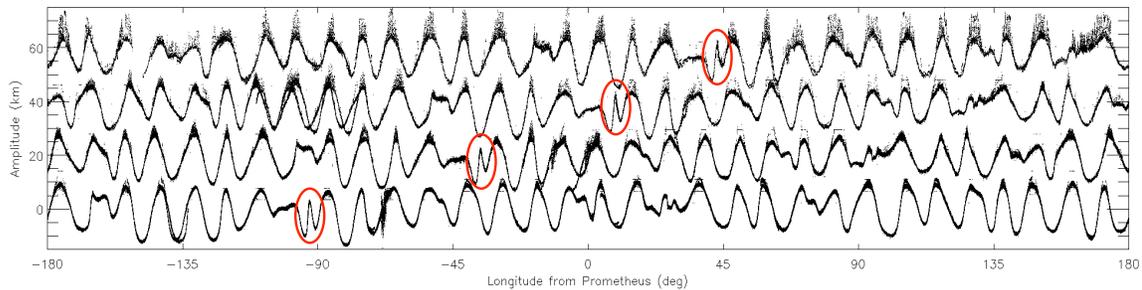

**Figure 7.** Mosaics of edge profiles taken in August 2013, with just a few days of time between each successive profile (Table 1). The longitude is relative to Prometheus. Although the m=32 pattern is following Prometheus, additional featrues (for example the circled one) move at the local Keplerian rate (see text).

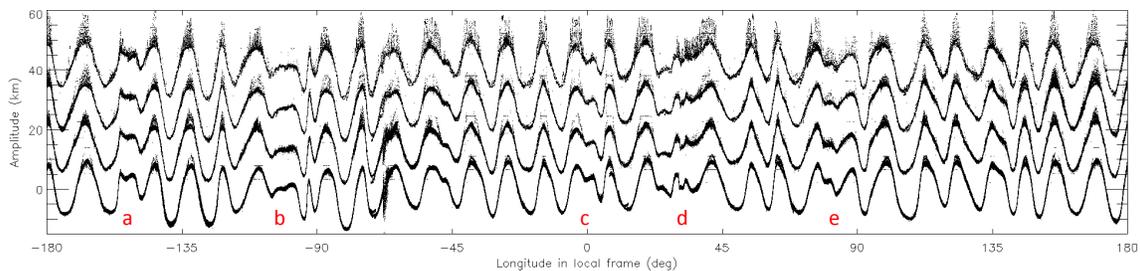

**Figure 8.** Mosaics of edge profiles taken in August 2013 (same ones in Fig. 7), with just a few days of time between sucessive profiles (see Table 1). The longitude is relative to a ring edge particle orbiting at a local Keplerian rate. The feature highlighted in Fig. 7 is here labeled as 'b'.

It might reasonably be asked if all pairs of nearby observations would be expected to match as closely as those in Fig. 8. The answer is "no", for two reasons. Firstly, all four FMOVIEs in Figs 7 and 8 were targeted at almost the same inertial longitude $\lambda$, within a range of 239 to 285°. As shown by El Moutamid et al. (2016), an *m*-lobed resonant perturbation with pattern speed $\Omega_p$, if viewed in the local keplerian frame, is described by the approximate expression

$$r(\theta) \approx a\{1 + e\cos[(m-1)\theta + \lambda + \delta]\} \qquad (6)$$

where $\theta = \lambda - n\,(t-t_0)$, $\lambda$ is the inertial longitude of the measurements and $\delta$ is the phase as defined in Eq. (2) above. This implies not only that the pattern has *m*-1 lobes in the local keplerian frame, as noted in Section 2 above, but also that the phase of the pattern depends on $\lambda$ as well as $\delta$. Only when $\lambda$ is similar for two mosaics can we expect to see similar phases for the resonant patterns in the local keplerian frame. (Another way to express this is to require that the two observations of a given ring segment be separated by an integral number of orbits.) The second limitation is imposed by apsidal precession. After more than a few weeks, the pericenters associated with any free eccentricity (ie., those not forced by the satellite resonance) will have precessed significantly, After one-half of the apsidal precession period (or ~60 days at the Keeler gap), what was initially a minimum in radius will appear as a maximum at the same inertial longitude, and the sign of the small-scale perturbations will be reversed.

In order to better understand what is going on here, it is useful to keep in mind four distinct dynamical time scales. First is the local keplerian orbital period, $P_{orb}$ = 14.3 hr. Second is the time it takes a ring particle at the inner edge of the Keeler gap to gain one 'lap' on Prometheus, i.e, the synodic period $P_{syn}$ = 19.26 day. In this time, the ring particle completes 32 orbits of Saturn and moves through all 32 lobes in the resonant pattern, each minimum corresponding to a pericenter passage in its saturnicentric orbit. (In this same period, of course, Prometheus completes 31 orbits around the planet.) Third is the local apsidal precession period, $P_{apse} = 2\pi/\dot{\varpi}$ = 120 day. Fourth is the synodic period of the ring particle relative to the satellite Daphnis, whose mean motion is 605.98 °/day, or $P_{enc}$ ~ 7.5 yr. This is the interval between successive close encounters with the nearest massive body, at which time one would expect any free eccentricity to be reset if it has not already been damped by collisions. The set of observations in Figs. 7 and 8 span a period which is < $P_{syn}$ and << $P_{apse}$ or $P_{enc}$. We thus expect to see minimal effects due to precession and no intervening Daphnis encounters. (Daphnis falls near -70° longitude in Fig. 8, as may be seen from the enhanced `noise' in our measured edge radii in this vicinity.)

At least two possible explanations seem possible for the persistent small-scale structure in Figs. 7 and 8. One is that there are relatively-large objects embedded in the ring, which orbit at the local keplerian rate and "carry with them" local radial distortions in the ring streamlines, at least until the next encounter with Daphnis. The second is that some groups of ring particles have slightly different eccentricities and/or pericenter longitudes than are predicted to arise from the 32:31 resonance with Prometheus and from the normal modes identified above. In the former case, a positive radial perturbation might be expected to remain positive at all longitudes, while in the latter case a minimum in radius will become a maximum one-half an orbit later.

We test these predictions using another set of mosaics made from observations with a different kind of geometry. As described in Section 1, FMOVIE mosaics usually involve the camera staring at a fixed inertial longitude and letting the ring particles pass through the camera's field of view. This was the case for the mosaics in Fig. 7 and 8. But in several instances, such as on revs 29, 173, 177, 183 198, 201 and 203, Cassini stared at one ring ansa (ie., at a fixed inertial longitude) for approximately half an orbital period, and then observed the *same ring particles* on the opposite ansa (~180° away in inertial longitude). Figure 9 shows edge profile mosaics for such an observation in September 2006, on rev 29. Here we see that the bottom profile (from movie 029a in Table 1) is almost the reverse of the middle profile (from movie 029b in Table 1). This is exactly what is expected for the *m*=32 resonant perturbation, since it forces the eccentricities and pericenters of ring particles rather than their semimajor axes: if a ring particle is at periapse at one ansa of the ring, it will be at apoapse 180° later. However, in Fig. 9 we see that this phenomenon also applies to the additional, small-scale perturbations. As the top profile in Fig. 9 shows, when we overlay the reverse of the middle profile (in red) over the bottom one (in black) the agreement is almost perfect. Clearly virtually all the patterns in both profiles are anti-aligned one-half an orbit later. This strongly suggests that most of the small-scale features are also due to eccentricity and/or pericenter variations, rather than to persistent radial perturbations such as an embedded object might produce. (An exception is the region between 130° and 140° where an embedded object is a possibility.) We have examined several such pairs of observations and in every case we find the same situation as in Fig. 9: clear evidence for eccentricity and/or pericenter variations but very little sign of persistent radial perturbations.

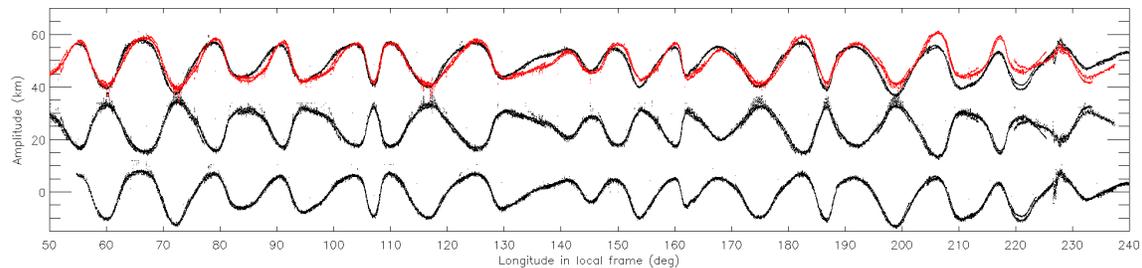

**Figure 9.** Mosaics of edge profiles obtained in September 2006, or rev 29 (see Table 1). The bottom profile was made when Cassini observed an inertial longitude of 266° as the ring particles passed through its field of view. The middle plot was made 23 hrs later, with Cassini observing the ring at an inertial longitude of 84°, as the same ring particles passed through the opposite ansa. The top profile is an overlay of the bottom and the reverse of the middle ones. The longitudes are measured relative to an arbitrary ring particle orbiting at the local keplerian rate.

What might be responsible for such azimuthal variations in the resonantly-forced eccentricity and/or pericenter? One possibility is that periodic encounters with Daphnis are able to perturb the eccentricities of ring particles. In this situation, the ring particles gain free eccentricities after encountering the satellite that damp several orbits later. (The visible manifestation of this is the wavy edges of the gap seen within a few degrees downstream from the satellite; see Fig. 1 for examples). A second possibility is also related to encounters with Daphnis, but in this case we focus on the much smaller perturbations in semi-major axis (which are second-order in the satellite-to-planet mass ratio; Dermott 1984). Over time, this will lead to some ring particles moving slightly forward relative to their slower companions ahead or behind them in longitude. As a result, initially sinusoidal variations in radius due to the 32:31 resonance can become asymmetric, and their wavelength reduced or increased locally, as we see in Figs. 7 and 8. However, it seems unlikely that this second process can lead to additional minima or maxima, such as those also seen in the same figures. Yet another possibility is that ring embedded objects on eccentric orbits near the inner edge of the Keeler gap could produce the observed localized features; the plausibility of this idea could be tested via simulations of interactions between such an eccentric embedded objects with the ring edge in order to verify if the perturbation reverses when the object is at periapse.

We have attempted to track these irregular features in all the available FMOVIEs, which span more then nine years in total. In Figure 10 we show a set of edge profiles for the inner edge of the Keeler gap, generated in the local keplerian frame as in Fig. 8, obtained between 2012 and 2015. Many features can be tracked here but we focus on four. Note that some of these features appear to be shaped differently, or even reversed in sign, compared to neighboring mosaics. This can happen for two reasons: firstly because Cassini observed the ring at different inertial longitudes in the two observations, so that the ring particles have moved to a different orbital phase (or that the pattern is reversed on the opposite ansa as in Fig. 9). The second reason is the secular apsidal precession of particle orbits noted above; even if Cassini always observed the same inertial longitude, precession will slowly move this from periapse to apoapse at a rate of ~2.97°/day. In either case, this is consistent with most of these irregularities being due to eccentricity/pericenter perturbations rather than to changes in semimajor axis.

We see from Fig. 10 that these small-scale features typically survive for up to a year before they are damped out. If such perturbations were related to embedded objects, this could mean that the objects are disrupted or change their orbits. (Tracking features for longer periods is frustrated by the fact that Cassini spent most of 2005 and 2006, as well as all of 2010 and 2011, in equatorial orbits, where ring observations were impossible.) We also note that the feature labeled A in Fig. 10 appears to have survived an encounter with Daphnis in about July 2013, suggesting that encounters with this nearby satellite do not erase or reset the perturbations. None of the other labeled features encountered Daphnis during the period of intensive Cassini observations in 2012 - 2014 (recall that $P_{enc}$ ~ 7.5 yrs). Similar comparisons of mosaics obtained in 2006/7 and in 2008/9 show further examples of persistent small-scale structures, but the observations are less extensive than those in Fig. 10.

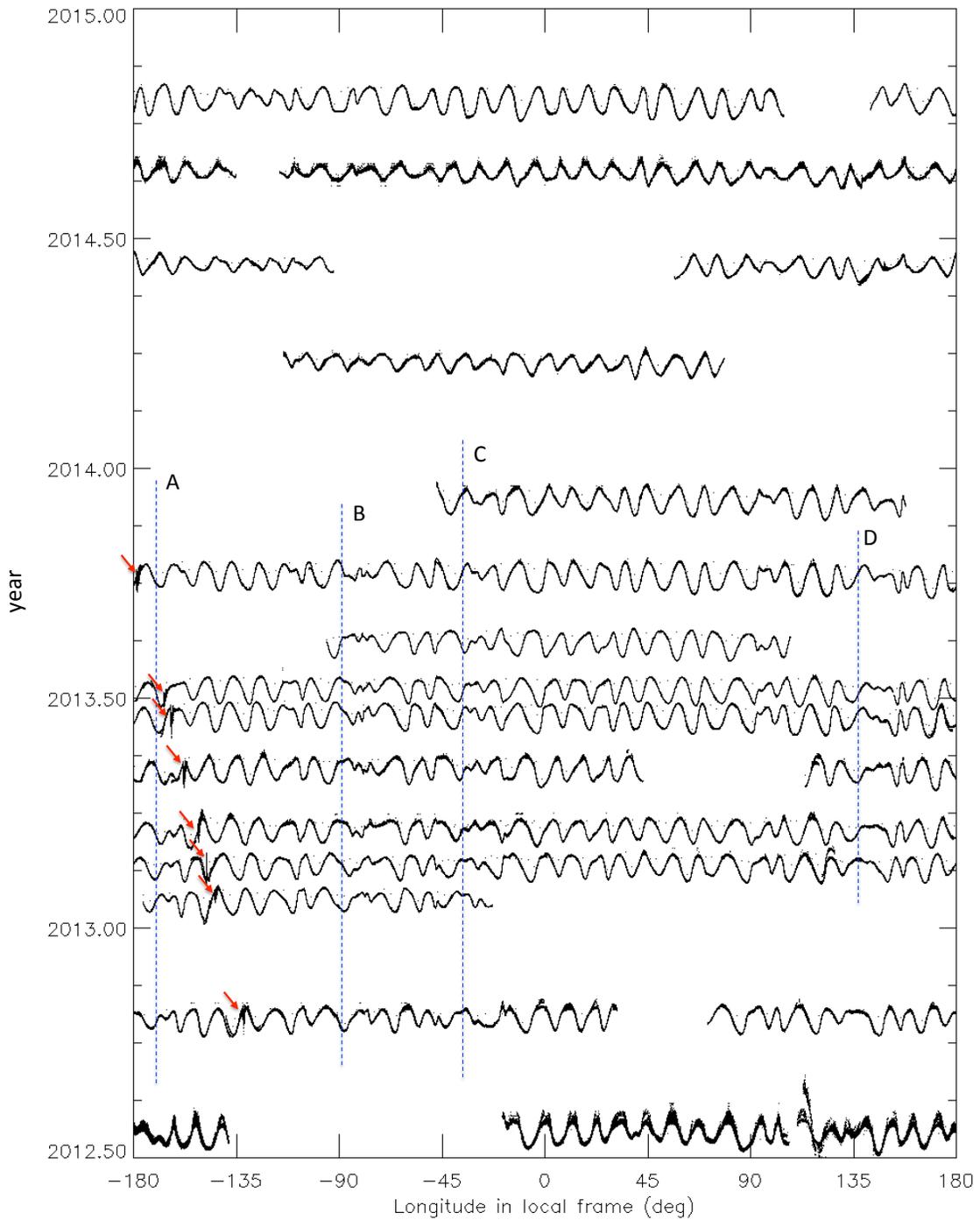

**Figure 10**. Profiles of the inner edge of the Keeler gap obtained between 2012 and 2014, assembled in the same local keplerian frame as Fig. 8. Several features of interest are located immediately to the right of the dashed lines labeled A, B, C and D. The red arrow points at the Daphnis wave crossing feature A.

## 6. Phase lag and torques

By analogy with the Mimas 2:1 ILR at the outer edge of the B ring and the Janus 7:6 ILR at the outer edge of the A ring (Spitale & Porco 2009, Spitale & Porco 2010, El Moutamid et al. 2016), one might expect that the Prometheus 32:31 ILR is responsible for confining the inner edge of the Keeler Gap. In such a situation, the torque $T_r$ exerted on the ring by the resonance should balance the viscous torque $T_v$, due to the radial transport of angular momentum through the ring. As shown by Borderies et al. (1982), the resonant torque is proportional to a phase lag in the ring's response, as well as to the square of the satellite's mass, while the viscous torque is proportional to the ring's effective kinematic viscosity ν, leading to the following approximate relation:

$$\sin m\Delta = \frac{\nu}{n(ma)^2}\left(\frac{M_p}{M_s}\right)^2 \qquad (7)$$

where $M_p$ and $M_s$ are the masses of the planet (Saturn) and the perturbing satellite (Prometheus), and $n$ is the mean motion of a ring particle at the edge of the gap (given by Eq. 4). In terms of our notation, $m\Delta = \delta - \pi$, which we found in Section 2 to be approximately -14°, albeit with considerable scatter. (A similar lag was observed in the $m=2$ pattern at the edge of the B-ring by Spitale & Porco (2010), and interpreted by these authors in terms of the B ring's effective viscosity ν.)

However, applying Eq. (7) to the inner edge of the Keeler gap, we obtain a very small viscosity of $\nu \approx 0.44$ cm$^2$/s. This number is almost three orders of magnitude lower than the viscosity of ~200 cm$^2$/s obtained from an analysis of the damping of density waves in nearby regions (Tiscareno et al. 2007). In fact, no solution of this equation is possible for a viscosity comparable to that obtained by Tiscareno et al. (2007).

Indeed, the torque from the 32:31 resonance with Prometheus is not the only torque acting to hold the Keeler gap open; Daphnis is expected to exert a shepherding torque on *both* edges that prevents them from viscously spreading and eventually closing the gap. We must therefore balance the viscous torque at the inner edge of the gap with the sum of the resonant torque from Prometheus and the shepherding torque from Daphnis. The three torques involved may be expressed as follows:

the resonant torque (Borderies et al. 1982):

$$T_r \approx (man)^2 e \frac{M_s}{M_p} \Sigma a \Delta a |\sin m\Delta|, \qquad (8)$$

the shepherding torque (Goldreich & Tremaine, 1982):

$$T_s \approx 0.84 a^4 \Sigma n^2 \left(\frac{a}{a_s - a}\right)^3 \left(\frac{M_s}{M_p}\right)^2, \qquad (9)$$

and finally the viscous torque (Borderies et al. 1984):

$$T_v = 3\pi\nu\Sigma n a^2, \qquad (10)$$

where the Σ and $v$ are the mean surface density and viscosity of the ring, respectively, and $ae$ is the radial amplitude of the perturbed edge, $\Delta a = ae$ and $a_s - a$ is the radial separation between Daphnis and the mean ring edge. The mass $M_s$ in Eq. (9) represents that of Daphnis since it is the shepherding satellite. Using a mass of Daphnis of $6.8 \times 10^{13}$ kg (Weiss et al., 2009, measured from the amplitude of the waves produced by Daphnis on the gap edges), and a surface density of 15 g/cm$^2$ (Tiscareno & Harris 2015, 2017), we estimate the sizes of these torques to be $T_v = 6.5 \times 10^{13}$ Nm (if we use a ring viscosity of 200 cm$^2$/s; Tiscareno et al. 2007), $T_s = 2.98 \times 10^{12}$ Nm, and $T_r = 5.9 \times 10^{11}$ Nm.

Since the resonant torque is almost two orders of magnitude weaker than the shepherding torque, we conclude that, regardless of the large uncertainty in the mean phase lag, the shepherding effect of Daphnis must dominate the maintenance of the edge, while the resonance with Prometheus plays a much smaller role.

Furthermore, the nominal viscous torque is much larger than the resonant and the shepherding torques put together, by a factor of almost 20. Assuming that the edge is not in fact governed by an unknown effect that is even larger than those under consideration, the simplest solution is to adjust the viscosity that enters into Equation 10. By balancing all three torques acting on the inner edge of the Keeler gap, $T_v = T_s + T_r$, we obtain an effective viscosity of $v \approx 11$ cm$^2$/s, more than an order of magnitude less than that inferred from density waves in the outer A ring (Tiscareno et al. 2007). We believe this result is an instructive warning against assuming that a viscosity calculated from one process can necessarily be applied to another process. Like the "α-parameter" in protoplanetary disk models, the viscosity is useful in calculations as a stand-in for more complex physics and can mean different things in different contexts. Although both processes no doubt have the similarity of being characterized by friction among ring particles, there is no reason to presume that the damping of spiral density waves and the spreading of a ring at its edges are identical processes in all particulars. Thus, we suggest that the value calculated here of $v \approx 11$ cm$^2$/s be taken alongside the value of $v \approx 200$ cm$^2$/s calculated by Tiscareno et al. (2007) as a basis for further work in understanding the details of dissipative ring processes.

## 7. Conclusion

Mosaics of the inner edge of the Keeler gap reveal a complex structure indicative of multiple dynamical phenomena. After processing several tens of thousands of Cassini images, grouped into 58 distinct FMOVIE observations, we find consistent evidence of a 32-lobed resonant perturbation with a mean amplitude of ~4.5 km. Phase analysis confirms that this pattern is following Prometheus' mean motion, with a small phase lag, and is thus almost certainly caused by the 32:31 inner Lindblad resonance with that satellite.

However, irregularities in the $m=32$ pattern suggest the presence of additional perturbations. Fourier analysis provides evidence for two additional long-wavelength perturbations, with $m=18$ and $m=20$. The former might be associated with the 18:17 inner

Lindblad resonance with Pandora, although the resonant radius is ~28 km interior to the ring edge. However, the best-fitting pattern speed is 0.155°/day slower than the mean motion of Pandora, suggesting a normal mode with a resonant radius which is only ~4 km interior to the ring edge. We hypothesize that the density wave driven by the 18:17 resonance indirectly excites the normal mode near the edge of the gap. The $m=20$ pattern, on the other hand, does not seem to be related to any satellite resonance, but is also consistent with a normal mode excited at the gap edge. Curiously, the resonant radius is the same as that of the $m=18$ mode. Both $m=18$ and $m=20$ perturbations have amplitudes of ~1-2 km.

In addition to the resonant and normal mode perturbations, we see several singular, small-scale features that appear to be orbiting Saturn at the local Keplerian rate. At any given time, 4 or 5 such features seem to be present on the ring edge, and to persist for periods of up to a year. Using observations which combine data from two opposite ring ansae, or inertial longitudes which differ by ~180º, we conclude that almost all of these features represent local eccentricity and/or pericenter perturbations rather than persistent radial distortions. Embedded objects at the ring edge could be the origin of those features, as proposed for similar features seen at the B ring's outer edge by Spitale and Porco (2010). An alternative hypothesis is that ring particles in different segments of the ring edge react slightly differently to the resonant perturbations from Prometheus, resulting in local variations in the forced eccentricities and/or pericenter longitudes. Equivalently, we may say that these regions develop small free eccentricities, which add to the global forced eccentricity due to the resonance. A similar free $m=2$ mode is observed at the edge of the B ring, of amplitude comparable to the resonantly-forced perturbation due to Mimas, but this appears to be global free mode rather than several local perturbations (Spitale & Porco 2010, Nicholson et al. 2014a).

Similar small-scale features appear at the outer edge of the A ring (Spitale & Porco 2009, El Moutamid et al. 2016), as well as that of the B ring (Spitale & Porco 2010, Nicholson et al. 2014a). We would encourage future examinations of paired observations of these edges obtained at opposite ring ansae, where available, in order to ascertain if these features also represent eccentricity/pericenter variations.

| Cassini orbit | Observation mid-time (UTC) | Observation duration (hours) | Inertial longitude (degree) | Mean radial resolution (km/pixel) | Number of images |
|---|---|---|---|---|---|
| 007 | 2005 MAY 01 17:46:30 | 13.7 | 224 – 232 | 3.4 | 247 |
| 008 | 2005 MAY 18 20:08:23 | 13.3 | 215 – 219 | 4.7 | 194 |
| 029_a | 2006 SEP 29 10:17:00 | 13.7 | 266 | 5.0 | 93 |
| 029_b | 2006 SEP 30 09:02:39 | 8.5 | 84 | 5.8 | 54 |
| 032 | 2006 NOV 14 00:23:35 | 30.5 | 94 – 104 | 5.0 | 130 |
| **036_1** | 2006 DEC 24 00:28:17 | **15.6** | **285 – 290** | **6.0** | **127** |
| **036_2** | 2007 JAN 06 03:53:58 | **13.4** | **255 – 260** | **5.3** | **130** |
| **039** | 2007 FEB 27 22:58:22 | **15.8** | **197 – 209** | **5.2** | **144** |
| **041_1** | 2007 MAR 31 22:07:40 | **12.5** | **175 – 179** | **6.0** | **129** |
| **041_2** | 2007 MAR 17 18:55:11 | **16.8** | **209 – 216** | **5.4** | **169** |
| **044** | 2007 MAY 05 15:17:16 | **14.1** | **178** | **6.4** | **134** |
| **055** | 2008 JAN 01 05:16:28 | **13.5** | **168 – 173** | **4.9** | **148** |
| **057** | 2008 JAN 24 03:20:07 | **13.2** | **143** | **5.1** | **132** |
| **083** | 2008 AUG 31 05:12:34 | **12.9** | **113** | **3.6** | **222** |
| **087** | 2008 OCT 01 03:53:54 | **11.4** | **284** | **2.7** | **171** |
| **089** | 2008 OCT 15 10:34:26 | **11.9** | **283** | **3.1** | **176** |
| **093** | 2008 NOV 15 00:30:25 | **9.4** | **126 – 132** | **3.2** | **200** |
| **100** | 2009 JAN 11 21:53:15 | **11.2** | **125 – 134** | **3.1** | **212** |
| **103** | 2009 FEB 11 00:51:45 | **10.4** | **167 – 193** | **3.2** | **213** |
| **105_2** | 2009 MAR 12 01:16:43 | **13.4** | **147 – 0** | **2.7** | **210** |
| **106** | 2009 MAR 24 00:00:22 | **12.9** | **146 – 190** | **2.8** | **208** |
| **108** | 2009 APR 16 20:26:15 | **10** | **237 – 247** | **4.0** | **210** |
| **110** | 2009 MAY 10 19:51:49 | **10.8** | **254 – 262** | **2.9** | **201** |
| **114** | 2009 JUL 14 08:10:56 | **12.2** | **223** | **4.7** | **130** |
| **115** | 2009 JUL 30 13:45:29 | **12.7** | **238** | **5.3** | **149** |
| **168** | 2012 JUN 25 12:03:40 | **9** | **283** | **5.7** | **76** |
| **172** | 2012 SEP 21 06:10:51 | **12.6** | **275 – 279** | **5.3** | **97** |
| **173_a** | 2012 OCT 16 03:41:39 | **8** | **140 – 110** | **4.1** | **55** |
| **173_b** | 2012 OCT 16 11:35:48 | **7.8** | **335 – 306** | **3.8** | **53** |
| **176** | 2012 DEC 07 09:23:52 | **14.9** | **-4 – 10** | **4.6** | **103** |
| **177_a** | 2012 DEC 19 02:22:45 | **5.8** | **239** | **4.9** | **51** |
| **177_b** | 2012 DEC 19 08:18:45 | **5.8** | **24** | **4.9** | **51** |
| **179** | 2013 JAN 14 09:18:38 | **15.1** | **261** | **5.0** | **143** |
| **180** | 2013 JAN 28 06:01:11 | **12.3** | **333** | **5.8** | **122** |
| **181** | 2013 FEB 10 09:13:55 | **15.8** | **270** | **4.9** | **140** |
| **183_a** | 2013 MAR 05 04:09:05 | **5.8** | **45** | **4.7** | **61** |
| **183_b** | 2013 MAR 05 10:09:05 | **5.8** | **190** | **5.6** | **61** |
| **184_2** | 2013 MAR 28 00:34:55 | **11.2** | **9** | **4.7** | **107** |
| **189** | 2013 MAY 07 02:36:31 | **14.4** | **239** | **4.4** | **140** |
| **191** | 2013 MAY 27 23:15:20 | **14.2** | **265** | **4.1** | **107** |
| **193** | 2013 JUN 20 14:47:57 | **15.8** | **45** | **4.2** | **150** |
| **194_a** | 2013 JUL 02 10:31:17 | **7.9** | **278** | **4.5** | **75** |
| **194_b** | 2013 JUL 02 18:29:07 | **7.9** | **111** | **5.6** | **75** |
| **196_3** | 2013 AUG 21 12:38:50 | **14.8** | **239** | **4.4** | **130** |
| **196_4** | 2013 AUG 25 00:52:50 | **14.8** | **278** | **5.7** | **130** |
| **196_5** | 2013 AUG 27 01:37:34 | **15.8** | **285** | **6.5** | **140** |
| **196_6** | 2013 AUG 28 23:53:20 | **14.3** | **300** | **7.2** | **125** |
| **197_2** | 2013 SEP 15 22:24:25 | **7.9** | **270** | **5.1** | **72** |
| **197_7** | 2013 SEP 07 18:59:43 | **10.6** | **255** | **4.8** | **92** |
| **198a** | 2013 OCT 19 00:50:25 | **7.9** | **278** | **6.0** | **70** |

| 198b  | 2013 OCT 19 08:48:15 | 7.9  | 110 | 6.0 | 70  |
|-------|----------------------|------|-----|-----|-----|
| 201_a | 2014 JAN 30 06:25:13 | 7.4  | 134 | 5.8 | 53  |
| 201_b | 2014 JAN 30 13:51:43 | 7.4  | 314 | 5.5 | 52  |
| 202   | 2014 MAR 02 11:01:53 | 12.8 | 148 | 6.6 | 95  |
| 203_a | 2014 APR 13 17:51:14 | 7.9  | 203 | 5.4 | 68  |
| 203_b | 2014 APR 14 01:49:14 | 7.9  | 35  | 5.7 | 70  |
| 205   | 2014 JUN 22 21:41:56 | 13.3 | 71  | 6.8 | 123 |
| 207   | 2014 AUG 16 23:29:52 | 12.5 | 233 | 4.1 | 110 |

**Table 1.** List of FMOVIE observations used in this work. The first column represents the FMOVIE ID; the first number represents the Cassini orbit followed by the sequence number within the same orbit; the letters a and b represent observations at opposite ring ansae.